\documentclass[prb,reprint,nofootinbib]{revtex4-1}
\bibpunct{[}{]}{;}{n}{}{}

\pdfoutput=1
\usepackage[caption=false]{subfig}

\usepackage{graphicx}
\usepackage{comment}
\usepackage{amsmath}
\usepackage{amssymb}
\usepackage{amsfonts}
\usepackage{bbm}
\usepackage{braket}
\usepackage{verbatim}
\usepackage[hidelinks]{hyperref}
\usepackage{ulem}
\usepackage{color, colortbl}
\usepackage{multirow}
\definecolor{LightCyan}{rgb}{1,0.5,0.5}
\hypersetup{
  colorlinks   = true, 
  urlcolor     = blue, 
  linkcolor    = blue, 
  citecolor   = blue 
}

\usepackage{color}
\usepackage{tikz}

\newcommand{\maxcut}{\texttt{MAXCUT} }

\newcommand{\be}{\begin{equation}}
\newcommand{\ee}{\end{equation}}

\begin{document}

	\title{Classically optimal variational quantum algorithms}
	
	\author{Jonathan Wurtz}
	\email[Corresponding author: ] {jonathan.wurtz@tufts.edu}
	\author{Peter Love}
	\affiliation{Department of Physics and Astronomy, Tufts University, Medford, Massachusetts 02155, USA}

	\date{\today}
	
	\begin{abstract}
    Hybrid quantum-classical algorithms, such as variational quantum algorithms (VQA), are suitable for implementation on NISQ computers.  In this Letter we expand an implicit step of VQAs: the classical pre-computation subroutine which can non-trivially use classical algorithms to simplify, transform, or specify problem instance-specific variational quantum circuits. In VQA there is a trade-off between quality of solution and difficulty of circuit construction and optimization. In one extreme, we find VQA for \texttt{MAXCUT} which are exact, but circuit design or variational optimization is \texttt{NP-HARD}. At the other extreme are low depth VQA, such as QAOA, with tractable circuit construction and optimization but poor approximation ratios. Combining these two we define the Spanning Tree QAOA (ST-QAOA) to solve \texttt{MAXCUT}, which uses an ansatz whose structure is derived from an approximate classical solution and achieves the same performance guarantee as the classical algorithm and hence can outperform QAOA at low depth. In general, we propose integrating these classical pre-computation subroutines into VQA to improve heuristic or guaranteed performance.
	\end{abstract}

	\maketitle
\begin{figure}
        \centering
        \includegraphics{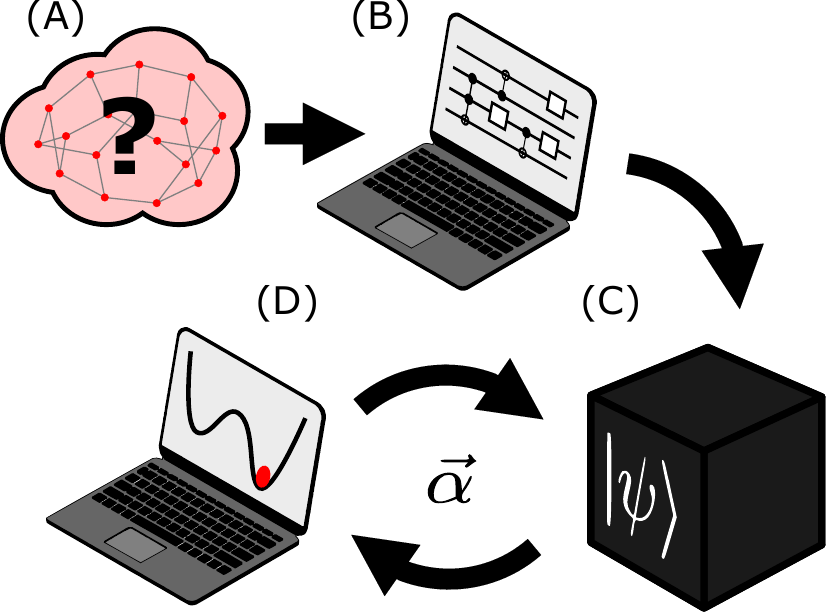}
        \caption{A pictorial representation of a variational quantum algorithm. Given some problem instance \textbf{(A)}, a classical subroutine \textbf{(B)} tailors a problem instance-specific circuit, by including problem structure, simplifying or transforming the problem, or using results and structure of classical algorithms. The circuit is run by a near term quantum machine \textbf{(C)} and variational parameters are optimized via repeated query of a classical optimizer \textbf{(D)}.}
        \label{fig:VQA_diagram}
    \end{figure}

	 Today's noisy intermediate scale quantum computers (NISQ) are bounded in power by size, noise and decoherence~\cite{Preskill2018}. Do there exist implementable hybrid quantum-classical algorithms  which outperform the best classical algorithms? Such an algorithm would exhibit quantum advantage, perhaps the most ambitious goal of the NISQ era. One class of algorithms which shows promise are Variational Quantum Algorithms (VQA) \cite{Peruzzo2014,farhi2014,McClean_2016,Moll_2018,cerezo2020variational}, which variationally optimize ans\"atze wavefunctions to extremize expectation values of objective functions. VQA construct a parameterized quantum circuit $U(\vec\alpha)$ in a classical pre-computation step (Fig.~\ref{fig:VQA_diagram}B), which is implemented on a NISQ device and optimized in an outer classical loop (Fig.~\ref{fig:VQA_diagram}C-D).

    Classical no-free-lunch (NFL) theorems for optimization imply that algorithmic advantages rely on problem structure~\cite{Wolpert1997}. Quantum NFL theorems for specific cases exist~\cite{poland2020no,sharma2020reformulation}, and for VQAs suggest that the ansatz should reflect problem structure, otherwise VQAs suffer from barren plateaus~\cite{mcclean2018barren}. How can the structure of the problem, form of classical algorithms, or approximate solutions, be used in the classical subroutines of VQA?

    A problem might be simplified or reduced as in~\cite{Bravyi2016}, or by exactly solving weakly connected parts of a \maxcut graph as in~\cite{Grotschel1984}. Alternatively, a problem might be mapped or transformed to one with better heuristics \cite{Peng2020,harrow2020small}. Classical insight may motivate the circuit structure as in the case of machine learning models~\cite{Benedetti_2019,Schuld2019}, or generate an analogous quantum version of a classical algorithm~\cite{brandao2017quantum}. 
    
    The coupled-cluster ansatz used in VQE~\cite{Peruzzo2014} and the QAOA ansatz~\cite{farhi2014} reflect structure by including the terms of the objective function in the ansatz. Additionally, VQAs can use approximate classical solutions in their ansatz states using the concept of warm starts, which initialize the variational parameters with values known to mimic a good classical solution~\cite{Gondzio1998,farhi2017_q,egger2020warmstarting,tate2020bridging}. If further variational optimization is possible, the VQA will improve upon the performance of the classical algorithm. In the worst case improvement may not be possible due to complexity theoretic constraints~\cite{haastad2001some,berman1999some,khot2007optimal,khotUG}.

     In this Letter, the warm start concept is generalized to construct problem instance-specific \textit{circuits} instead of just choosing initial variational parameters that reproduce approximate solutions in a fixed ansatz circuit structure. We use the VQA pre-computation step (Fig.~\ref{fig:VQA_diagram}B) to generate problem instance-specific circuits that use problem structure from the form and solutions of classical algorithms as well as from the objective function.

    While this pre-computation step is general, we focus on solving the particular problem of \maxcut inspired by a particular classical algorithm, with the hope of inspiring other VQA algorithms with non-trivial pre-computation subroutines. We construct the Spanning Tree QAOA (ST-QAOA), a particular VQA to solve \texttt{MAXCUT} \cite{GAREY1976,Hstad2001}. This algorithm uses approximate solutions from a classical \maxcut solver as a subroutine to construct a problem instance-specific circuit with $r$ rounds of gates. We show that $r=1$ is guaranteed to match the performance of the classical solver, and $r\to \infty$ approaches the exact result.

    We introduce the ST-QAOA in a sequence of algorithms that illustrate the trade-offs between quality of solution and classical computational difficulty of the pre-computation step. First, we introduce the Spanning Tree Algorithm, which can produce exact solutions at the expense that circuit generation is \texttt{NP-HARD}. Next, we introduce the Variational Spanning Tree Algorithm, which can produce optimal answers at the expense that variational optimization is \texttt{NP-HARD}. Finally, we introduce the ST-QAOA and present numerical evidence of its performance on random instances of $3$-regular \texttt{MAXCUT}, demonstrating that ST-QAOA can always reproduces the performance of the classical algorithm it uses to construct its ansatz.

    First, let us define the \texttt{SIGNED MAXCUT} problem and the structure of spanning trees, which will be the algorithmic insight for ST-QAOA. A signed graph $\Gamma = (\mathcal G,\sigma)$ \cite{Zaslavsky2018} is constructed of graph $\mathcal G$ and signature for each edge $\sigma=\pm 1$. The goal of \texttt{SIGNED MAXCUT} is to find a bipartition of vertices $\{X,Y\}$  (or binary string $z$ labeling the bipartition) such that the maximal number of edges of $\Gamma$ are satisfied (or ``cut"). An edge with a negative signature is satisfied if its vertices are in opposite partitions, and unsatisfied otherwise. \texttt{MAXCUT} is the specific case where the signature of every edge is negative.
    
    \texttt{SIGNED MAXCUT} is closely related to balance in signed graphs \cite{ZASLAVSKY1982}. A signed graph is balanced (``bipartite") if there exists a bipartition of vertices such that every edge of $\Gamma$ is satisfied \cite{harary1953}. \texttt{SIGNED MAXCUT} is equivalent to the maximum balanced subgraph problem \cite{Poljak1993}: given some signed graph $\Gamma$, what is the minimal number of edges $e$ which need be removed to make $\Gamma \backslash e$ balanced? Any solution $z$ induces some subset of edges $e$ which remain unsatisfied, so that $\Gamma\backslash e$ is balanced. An optimal solution $z$ will remove the smallest number of unsatisfied edges $e$.
    
    The bipartition $z$ of a balanced graph can be found with a directed spanning tree, as follows \cite{HARARY1980}. Given some balanced graph $\Gamma\backslash e$, construct any spanning tree $\mathcal T$ with a unique path between each vertex. Starting with some arbitrary origin vertex, traverse the tree to assign each vertex to a bipartition. If the signature of an edge is ($-$), assign the next vertex in the path to the opposite partition as its parent, and the same if ($+$). This satisfies every edge of the spanning tree and, because the graph is balanced, every edge in the reduced balanced graph $\Gamma\backslash e$. Any spanning tree over a balanced graph generates the satisfying bipartition $z$, as $z$ is unique \cite{zaslavsky2018negative}. Thus, the maximal bipartition $z$ of graph $\Gamma$ is equivalent to some choice of spanning tree $\mathcal T_z$. In this way, the search space of $\texttt{SIGNED MAXCUT}$ solutions can be reduced to searching through the set of all possible spanning trees \cite{POLJAK1986}, as the optimal bipartition $z$ is given by some particular spanning tree(s) $\mathcal T_z$ over the signed graph $\Gamma$.

    Given some signed graph $\Gamma = (\mathcal G,\sigma)$, the optimal bipartition of vertices is encoded in the maximal eigenstate of the objective function
    
    \begin{equation}
        C = \frac{1}{2}\sum_{\langle ij\rangle} (1+\sigma_{ij}\sigma_z^i\sigma_z^j),
    \end{equation}
    where each clause in the sum represents an edge of the graph, with eigenvalue $+1$ if the edge is satisfied and eigenvalue $0$ if the edge is not satisfied, and $\sigma_{ij}$ the signature of the edge. 
    
    The goal of any VQA is to optimize the expectation value of the objective function with an ansatz wavefunction. We write an ansatz circuit of $r$ rounds in the general form:
    \begin{equation}\label{eq:general_QAOA_ansatz}
        |\psi\rangle = e^{iH_1\alpha_1}e^{iH_2\alpha_2}(\cdots)e^{iH_r \alpha_r}|+\rangle.
    \end{equation}
    Note that $r$ does not correspond to circuit depth. The circuit is constructed from a restricted set of generators $H_q$; for example, in QAOA, $\sigma_z^i\sigma_z^j$ operators acting on all edges of a graph $\langle ij\rangle\in\mathcal G$, and $\sigma_x^i$ operators acting on all vertices. The classical computers' challenge is specifying the circuit structure in~(\ref{eq:general_QAOA_ansatz}) via the pre-computation step (Fig.~\ref{fig:VQA_diagram}B) and finding particular angles (Fig.~\ref{fig:VQA_diagram}D) which maximize the objective function. Optimal bitstring solutions may be read out by observing $\langle \psi|\sigma_z^i|\psi\rangle$ for each qubit and assigning the bitstring according to the measurement $\pm1$.
    
    We will now give exact VQA algorithms for \texttt{MAXCUT}. For the restricted generators of QAOA, eigenstates are cat states due to $\mathbb{Z}_2$ symmetry and the ansatz wavefunction lies in the +1 sector, so an optimal state must have the particular form of a ``cat state"~\cite{farhi2017_q}
    \begin{equation}\label{eq:Zcat_state}
        |\psi\rangle = \frac{1}{\sqrt 2}\big(|z\rangle + |\overline z\rangle\big),
    \end{equation}
    where $z$ is the binary representation of the optimal \texttt{MAXCUT} solution, and $\overline z=\neg z$. The ansatz of Eq.~\eqref{eq:general_QAOA_ansatz} can generate such states, and so VQA can be exact \cite{farhi2017_q}. Consider the sequence of unitaries between two vertices
    
    \begin{equation}\label{eq:generate_cat_state}
        |\psi\rangle = e^{i\pi\sigma_x^1/4}e^{i\pi\sigma_z^0\sigma_z^1/4}|++\rangle=\frac{1+i}{2}(|01\rangle + |10\rangle).
    \end{equation}
    
    This is the desired ``cat state", up to a global phase. Changing the sign of $\sigma_z\sigma_z$ generates the state $(|11\rangle + |00\rangle)/\sqrt{2}$. In general, a unitary $U_{ij}^{\pm}$ written as a directed arrow between two vertices
    
    \begin{gather}\label{eq:pm_unitary_define}
        U_{ij}^{\pm}=e^{i\pi\sigma_x^j/4}e^{\mp i\pi\sigma_z^i\sigma_z^j/4}\\\big\Updownarrow\nonumber\\
        \includegraphics{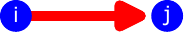}\nonumber
    \end{gather}
    evolves an initially unentangled X product state into a maximally entangled Bell pair with either the same ($+$) or opposite ($-$) correlation. It is simple to generalize that products of these unitaries along a directed tree will generate $Z$ cat states \cite{farhi2017_q}. For example,
    
    \begin{gather}
        U_{2,4}^-U_{1,2}^-U_{1,3}^+U_{0,1}^-|+\rangle = \frac{-1}{\sqrt{2}}(|01011\rangle + |10100\rangle).
        \\\big\Updownarrow\nonumber\\
        \includegraphics{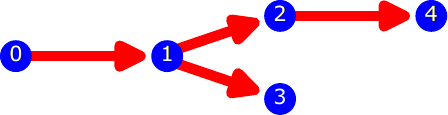}\nonumber
    \end{gather}

    If these unitaries map to a directed spanning tree of $\mathcal G$, they may prepare any eigenstate $z$ of Eq.~\eqref{eq:Zcat_state}, by choosing signs $U^\pm$ of each directed edge depending if the vertices are in the same (+) or opposite ($-$) partitions. These unitaries are Clifford: Eq.~\eqref{eq:generate_cat_state} is equivalent to a Hadamard on register 1, then a CNOT between registers 0 to 1.
    
    This is a unitary version of the spanning tree algorithm of~\cite{HARARY1980}, if one chooses the sign of each unitary to be the sign of the signature of its edge. Given some bipartition $z$ and associated spanning tree $\mathcal T_z$, one can construct the state as an ordered product of these unitaries over directed edges (up to a global phase)
    
    \begin{equation}\label{eq:exact_zzbar}
        \frac{|z\rangle + |\overline z\rangle}{\sqrt{2}} = \mathcal T\prod_{\langle ij\rangle\in \mathcal T_z}U_{ij}^{\sigma_{ij}}|+\rangle.
    \end{equation}
    
    Here, $\mathcal T$ denotes inverse path ordering of unitaries along the directed spanning tree $\mathcal T_z$, and $\sigma_{ij}$ is the signature of edge $\langle ij\rangle$ in signed graph $\Gamma$. Note that not every bipartition $z$ may be constructed in this manner, as the spanning tree requires the reduced balanced subgraph $\Gamma\backslash e$ to be connected. We call this algorithm ``Spanning Tree".
    
    This algorithm is exact in the following case. As part of the pre-computation step, some classical algorithm finds the optimal partition $z$ and an associated spanning tree $\mathcal T_z$. Then, implement the circuit of Eq.~\eqref{eq:exact_zzbar} to generate a maximal eigenstate of the objective function. However, this exactness comes at the cost that \textit{generating} the sequence of gates is classically hard. The classical algorithm which creates the optimal circuit must first find the spanning tree(s) whose bipartition provides the solution to the \texttt{SIGNED MAXCUT} problem, which is known to be $\texttt{NP-HARD}$ \footnote{\texttt{SIGNED MAXCUT} can be reduced to \maxcut by replacing every positive edge with two negative edges and a connectivity 2 vertex.}. This demonstrates a case in which their exist exact quantum circuits that provide solutions to hard problems, generating the circuit may itself be a hard problem classically.

    The Spanning Tree algorithm can be made variational by allowing the unitaries of the directed edges of any spanning tree to be a function of angles
    
    \begin{figure*}
        \centering
        \includegraphics[width=\linewidth]{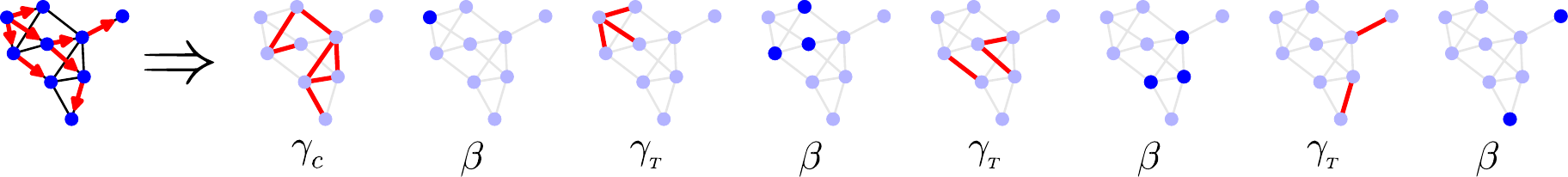}
        \caption{An example round of Spanning Tree QAOA. A classical algorithm computes some approximately optimal \texttt{SIGNED MAXCUT} solution by finding an appropriate spanning tree (left). A QAOA round alternates between $\sigma_z\sigma_z$ unitaries on edges (red) and $\sigma_x$ unitaries on vertices (blue). First, the complement unitary is applied, then the tree unitary, which mimics the directed tree graph via repeated application of $\sigma_z\sigma_z$ and $\sigma_x$. This is repeated $r$ times. For $r=1$ this is guaranteed to match the performance of the classical algorithm, and for $r\to\infty$ the approximation ratio approaches 1.}
        \label{fig:example_round}
    \end{figure*}

    \begin{equation}
        U_{ij}^\pm(\gamma,\beta) =e^{i\beta\sigma_x^j}e^{ \mp i\gamma\sigma_z^i\sigma_z^j}.
    \end{equation}
    
    The variational algorithm is as follows. For a signed graph $\Gamma$ of $N$ vertices, a classical algorithm generates a random spanning tree $\mathcal T$ and outputs a circuit which is a function of $2N-2$ angles
    
    \begin{equation}
        U(\vec \gamma,\vec \beta) = \mathcal T \prod_{q\in\langle ij\rangle\in\mathcal T_z}U_{ij}^+(\gamma_q,\beta_q),
    \end{equation}
    where $q$ indexes the edges of the tree, and $\mathcal T$ indicates the path ordering of unitaries along the randomly chosen directed spanning tree. We call this algorithm ``Variational Spanning Tree" (VST). 
    
    By Eq.~\eqref{eq:pm_unitary_define}, extremal values $\gamma_q\in\pm \pi/4$ and $\beta_q=\pi/4$ can construct any state $(|z\rangle + |\overline z\rangle)/\sqrt{2}$. It is the job of a classical optimizer to optimize the angles and find $N-1$ signs of $\gamma_p$ which construct the optimal state. However, this VST algorithm is classical. In the Heisenberg picture, expectation values of operators are
    
    \begin{align}\label{eq:simple_expectation_values}
        \langle \sigma_z^i\sigma_z^j\rangle = \prod_{q \in\text{ path $i\leftrightarrow j$}} \sin(2\gamma_q)\sin(2\beta_q),
    \end{align}
    where $q$ index all of the edges of the spanning tree on the unique path between vertices $i$ and $j$. Thus, the expectation value of the objective function for any graph may be computed classically. Bitstrings can also be efficiently sampled using tensor networks \cite{ferris2012} by recursively contracting leaves of the spanning tree. Hence VST is purely classical and cannot exhibit any quantum advantage. Instead of generating the circuit being \texttt{NP-HARD}, the optimization itself is \texttt{NP-HARD} \cite{Shaydulin2019,cerezo2020costfunctiondependent,bittel2021}. Ultimately, this is because the optimization algorithm is a discrete search, finding $N-1$ signs $\gamma=\pm \pi/4$ or analogously the optimal bitstring $z$. VST demonstrates that while circuit generation may be easy, optimizing parameters in and of itself may be a hard problem classically.

    It is reasonable to expect that if the number of variational parameters is constant in problem size, the optimization is more efficient. QAOA has this property~\cite{farhi2014} due to having a more constrained ansatz. Instead of choosing individual terms $\sigma_z\sigma_z$ and $\sigma_x$ acting in serial, QAOA alternates between acting with the objective function $C$ and a sum of Pauli $\sigma_x$ terms $B$. The number of variational angles is $2p$, independent of problem size
    \begin{equation}
        |\gamma,\beta\rangle = e^{-iB\beta_p}e^{-iC\gamma_p}(\cdots)e^{-iB\beta_1}e^{-iC\gamma_1}|+\rangle.
    \end{equation}
    
    There has been much work on QAOA. The approximation ratio uniformly increases in $p$, with the $p\to \infty$ limit converging to the exact state with an adiabatic schedule \cite{farhi2014,crooks2018performance}. For large $p$, it has been observed that the optimal parameters exhibit concentration and become independent of graph instance \cite{brandao2018}, and optimal parameters for $p$ can induce the parameters for $p+1$ \cite{Zhou_2020}. These facts suggest that the classical optimization may be efficient. This comes at the cost that performance guarantees are combinatorially difficult to compute~\cite{farhi2014,wurtz2020}, and to date QAOA has not outperformed the best classical algorithms~\cite{farhi2014,wurtz2020}. QAOA can be contrasted with the VQAs we define above, which can generate eigenstates of the objective function, including the maximal state, at expense of generating the gate set, or the optimization procedure, being \texttt{NP-HARD}.
    
    We now propose a combination of the Spanning Tree algorithm and QAOA, which we call Spanning Tree QAOA (ST-QAOA). To avoid the difficulty of parameter optimization, the circuit optimizer will use a non-extensive number of variational parameters, like QAOA. To avoid the difficulty of circuit design, instead of finding the exact maximal spanning tree, the circuit construction will use an \textit{approximate} solution from a classical \maxcut solver to generate spanning trees. Finally, to mimic QAOA, the ansatz will repeat $p$ times, and be able to reproduce QAOA as a special case. The ST-QAOA is as follows.
    
    Given some signed graph $\Gamma$, a classical \maxcut solver $P$ outputs some bitstring $P(\Gamma)=z$ and associated directed spanning tree $\mathcal T_z$ with root vertex $v$ \footnote{Any classical algorithm can find solutions where the reduced balanced subgraph $\Gamma\backslash e$ is connected and so will have an associated spanning tree by adding the following subroutine: for all vertices within some disconnected subgraph of $\Gamma\backslash e$, swap the bipartition $X\Leftrightarrow Y$. This will satisfy all of the edges of the original graph between the two previously disconnected subgraphs, increasing the number of satisfied edges and making the reduced balanced graph connected.}. Next, partition the edges of the graph $\mathcal G$ into those in the spanning tree and its complement to define a ST-QAOA round
    
    \begin{multline}
        U(\gamma_c,\gamma_T,\beta)\equiv\\ \bigg(\mathcal T \prod_{\langle ij\rangle\in\mathcal T_z}U_{ij}^{\sigma_{ij}}(\gamma_T,\beta)\bigg)e^{i\beta\sigma_x^v}\bigg(\prod_{\langle ij\rangle\not \in \mathcal T_z}e^{i\gamma_c\sigma_{ij}\sigma_z^i\sigma_z^j}\bigg).
    \end{multline}
    
    An example of this circuit is shown in Fig.~\ref{fig:example_round}. This applies a unitary generated by the complement edges of the spanning tree, then the spanning tree unitary with some global angle. A ST-QAOA procedure repeats this unitary $p$ times as a function of $3p$ variational parameters
    
    \begin{equation}
    |\psi\rangle = U_{\text{ST-QAOA}}|+\rangle = \prod_{q=1}^r  U(\gamma_c^q,\gamma_T^q,\beta^q)|+\rangle. 
    \end{equation}
    
    Like QAOA, this algorithm has a number of parameters independent of the problem size. Unlike QAOA, the ansatz depends on the approximate classical solution, and can generate extensive correlations even for $r=1$. Such a circuit includes all terms in the objective function and mixing term, except reordered to include the structure of spanning trees, which allows the algorithm to include QAOA as a special case. Let us now inspect the performance of this algorithm.
    
    For $r=1$, ST-QAOA can return the bitstring $z$ produced by the classical subroutine $P$. By choosing the angles $\gamma_c = 0$, $\gamma_T =\pi/4$, $\beta = \pi/4$, the unitary is equivalent to Eq.~\eqref{eq:exact_zzbar} for the particular choice of spanning tree generated by $P$, and so ST-QAOA can give the same solution as the classical subroutine. Therefore ST-QAOA has the same performance guarantee as its classical subroutine.
    
    For $r=2$, it is possible to reproduce a round of QAOA. Given QAOA angles $\gamma_*$ and $\beta_*$,  for the first round, choose angles $\gamma_c=\gamma_T=\gamma_*$, and $\beta=0$. For the second round, choose angles $\gamma_c=\gamma_T=0$ and $\beta=\beta_*$. The first round of unitaries is equivalent to the unitary generated by the objective function, as each $\sigma_z\sigma_z$ term commutes. The second round of unitaries is equivalent to the unitary generated by the mixing function for the same reason. Thus, the approximation ratio of level $2r$ ST-QAOA will always be \textit{at least} that of level $p$ QAOA. As $p$ increases for QAOA, the approximation ratio increases, approaching $1$ in the $p\to\infty$ limit \cite{farhi2014}. Because the ST-QAOA includes QAOA as a special case, ST-QAOA will also approach the exact result as $r\to\infty$.

    \begin{figure}
    \centering
    \includegraphics{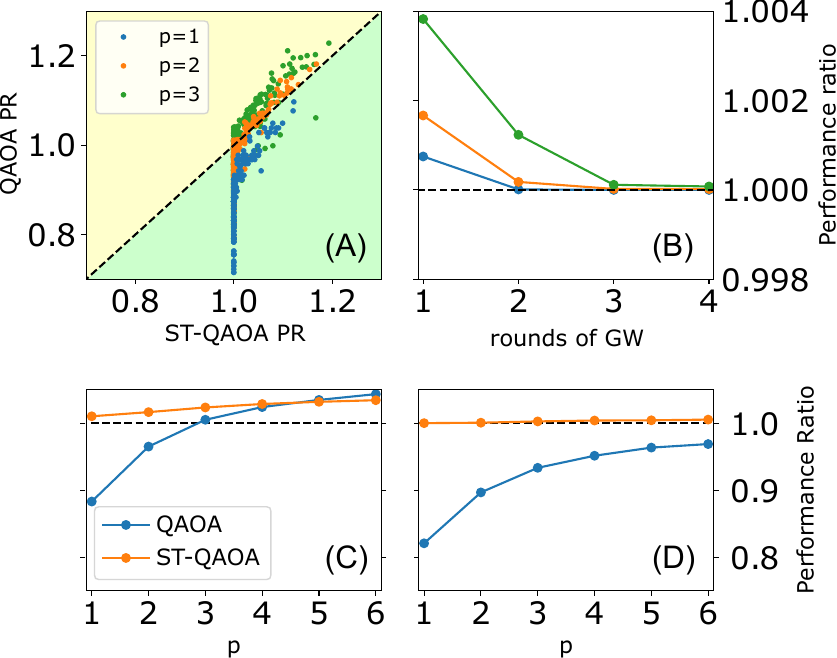}
    \caption{Comparing Spanning Tree QAOA (ST-QAOA), QAOA, Random Spanning Tree (RST) and Goemans-Williamson (GW) Algorithms for an ensemble of 250 random 3-regular graphs with 16 vertices. 
    \textbf{(A)} plots the performance ratio $B(\Gamma)$, comparing ST-QAOA (horizontal axis) and QAOA (vertical axis) for each graph in the ensemble. It is clear that the ST-QAOA has a performance guarantee $B(\Gamma)\geq 1$ for $p\geq 1$. 
    \textbf{(B)} plots the average performance ratio vs. the number of repetitions of the GW algorithm. As GW is a randomized algorithm, repeated sampling uniformly increases its performance, and the performance ratio appears to converge to 1 from above. 
    \textbf{(C,D)} plots the average performance ratio over the ensemble vs. the random spanning tree \textbf{(C)} and Goemans Williamson \textbf{(D)} algorithms. QAOA has advantage over the random algorithm (black dashed) with $p\geq3$ and advantage over ST-QAOA with the random algorithm for $p\geq 5$. 
    }\label{fig:goldenplot}
    \end{figure}
    
   ST-QAOA is a useful algorithm with which to interrogate the possibility of various forms of quantum advantage. Can ST-QAOA exceed the performance of the best classical algorithms for worst case graphs? This would be the case if further variational optimization is possible in ST-QAOA for all graphs when the best classical algorithm is used as a subroutine, and would represent quantum advantage. Such a case cannot be established numerically, and may not be possible due to complexity-theoretic constraints~\cite{haastad2001some,berman1999some,khot2007optimal,khotUG}. A simpler question is whether, above some threshold value of $r$, there exist subsets of graphs for which ST-QAOA has strictly better performance. This is not quantum advantage, as it only shows improved performance relative to a particular algorithm for a subset of graphs, which imposes additional structure that could be exploited by another specialized classical or quantum algorithm.
    
    To make a quantitative comparison of algorithms we use the performance ratio
    \begin{equation}
        B(\Gamma)=C_\text{Q}(\Gamma)\;/\;C_\text{C}(\Gamma),
    \end{equation}
    where $C_\text{Q}(\Gamma)$ is the optimized expectation value of the objective function for a VQA applied to the signed graph $\Gamma$, and $C_\text{C}(\Gamma)$ is the number of satisfied edges given an output from the competing classical algorithm $P$. A value $B(\Gamma)>1$ indicates that the quantum algorithm can find better solutions than the classical algorithm for particular problem instance $\Gamma$. If $\langle B\rangle>1$ for graphs in some ensemble $\{\Gamma\}$, then the quantum algorithm has average case quantum advantage over its classical competitor, as the quantum algorithm can produce better solutions than the classical algorithm in at least some of the graphs. For the ST-QAOA, the worst case $B(\Gamma)\geq 1$ for all graphs relative to the classical algorithm employed by ST-QAOA to generate the ansatz.
    
    As a numerical demonstration, we implement ST-QAOA on an ensemble of 3 regular graphs. We choose two classical algorithms to generate spanning trees. The first is that of Goemans and Williamson (GW) \cite{goemans1995}, which uses semidefinite programming to achieve an approximation ratio of at least $0.878$ in general graphs, and $0.932$ in $3$-regular graphs~\cite{halperin2004max}. The second algorithm samples a random spanning tree and achieves an approximation ratio of $2/3(1-1/n)$ for a $3$-regular graph with $n$-vertices~\cite{POLJAK1986}. Numerical results for an ensemble of $250$ $3$-regular graphs with $16$ vertices are shown in Fig.~\ref{fig:goldenplot}.  Optimization used gradient ascent initialized from $100$ random points in parameter space. From Fig.~\ref{fig:goldenplot}A, it is clear that $B(\Gamma)\geq1$ for ST-QAOA and that there exist graph instances for which the ST-QAOA exhibits advantage  over its competing classical algorithm and performs better than QAOA. Figs.~\ref{fig:goldenplot}C-D show that the average performance ratio $\langle B(\Gamma)\rangle\geq 1$ for all $p$ as expected, indicating an average case advantage for the spanning tree algorithm over its classical subroutine. Fig~\ref{fig:goldenplot}B shows that care must be taken to ensure that the classical algorithm is also performing optimally by illustrating the convergence of average performance ratio to one with increased number of rounds of GW. This illustrates that careless use of classical algorithms can create the illusion of quantum advantage.

    \quad
    
    In this Letter, we have expanded VQA with problem instance-specific circuits pre-computed by a classical subroutine. We also highlight the computational complexity pitfalls which may arise from such constructions. One may design an algorithm whose circuit yields exact answers to \texttt{MAXCUT}, at the cost that the algorithm is \texttt{NP-HARD}, as exemplified by the spanning tree algorithm. Similarly, an algorithm which constructs the circuit may be simple at the cost that the variational optimization algorithm is \texttt{NP-HARD}, as exemplified by the variational spanning tree algorithm. The intermediate algorithm, the Spanning Tree QAOA, combines QAOA with the concept of spanning trees to improve approximate classical solutions.
    
    However, the ST-QAOA requires the use of its competition as a subroutine to generate the circuit. In practice, any classical algorithm could integrate a similar scheme by running multiple algorithms in parallel and choosing the more optimal result, or use additional classical algorithms to improve the output of one classical algorithm. This is clear in Fig.~\ref{fig:goldenplot}B, when the optimal result among multiple GW queries is used as a classical solution. Because GW is randomized, the performance uniformly increases and removes any additional advantage from the ST-QAOA. Due to this subtlety, we make no claims of quantum advantage, even though the ST-QAOA can only increase the quality of solutions of its classical subroutine.
    
    While we focus on the problem \maxcut using the concept of spanning trees, the pre-computation step is more general. Using these ideas, constructing VQAs which take advantage of the pre-computation step to non-trivially generate problem instance-specific circuits may improve heuristic and guaranteed performance on the limited quantum resources of today's NISQ devices.

    \subsection*{Acknowledgements}
    This material is based upon work supported by the Defense Advanced Research Projects Agency (DARPA) under Contract No. HR001120C0068.
    
    \normalem
	\bibliographystyle{apsrev4-1}
	\bibliography{citationlist} 

\begin{thebibliography}{46}%
\makeatletter
\providecommand \@ifxundefined [1]{%
 \@ifx{#1\undefined}
}%
\providecommand \@ifnum [1]{%
 \ifnum #1\expandafter \@firstoftwo
 \else \expandafter \@secondoftwo
 \fi
}%
\providecommand \@ifx [1]{%
 \ifx #1\expandafter \@firstoftwo
 \else \expandafter \@secondoftwo
 \fi
}%
\providecommand \natexlab [1]{#1}%
\providecommand \enquote  [1]{``#1''}%
\providecommand \bibnamefont  [1]{#1}%
\providecommand \bibfnamefont [1]{#1}%
\providecommand \citenamefont [1]{#1}%
\providecommand \href@noop [0]{\@secondoftwo}%
\providecommand \href [0]{\begingroup \@sanitize@url \@href}%
\providecommand \@href[1]{\@@startlink{#1}\@@href}%
\providecommand \@@href[1]{\endgroup#1\@@endlink}%
\providecommand \@sanitize@url [0]{\catcode `\\12\catcode `\$12\catcode
  `\&12\catcode `\#12\catcode `\^12\catcode `\_12\catcode `\%12\relax}%
\providecommand \@@startlink[1]{}%
\providecommand \@@endlink[0]{}%
\providecommand \url  [0]{\begingroup\@sanitize@url \@url }%
\providecommand \@url [1]{\endgroup\@href {#1}{\urlprefix }}%
\providecommand \urlprefix  [0]{URL }%
\providecommand \Eprint [0]{\href }%
\providecommand \doibase [0]{http://dx.doi.org/}%
\providecommand \selectlanguage [0]{\@gobble}%
\providecommand \bibinfo  [0]{\@secondoftwo}%
\providecommand \bibfield  [0]{\@secondoftwo}%
\providecommand \translation [1]{[#1]}%
\providecommand \BibitemOpen [0]{}%
\providecommand \bibitemStop [0]{}%
\providecommand \bibitemNoStop [0]{.\EOS\space}%
\providecommand \EOS [0]{\spacefactor3000\relax}%
\providecommand \BibitemShut  [1]{\csname bibitem#1\endcsname}%
\let\auto@bib@innerbib\@empty
\bibitem [{\citenamefont {Preskill}(2018)}]{Preskill2018}%
  \BibitemOpen
  \bibfield  {author} {\bibinfo {author} {\bibfnamefont {J.}~\bibnamefont
  {Preskill}},\ }\href {\doibase 10.22331/q-2018-08-06-79} {\bibfield
  {journal} {\bibinfo  {journal} {Quantum}\ }\textbf {\bibinfo {volume} {2}},\
  \bibinfo {pages} {79} (\bibinfo {year} {2018})}\BibitemShut {NoStop}%
\bibitem [{\citenamefont {Peruzzo}\ \emph {et~al.}(2014)\citenamefont
  {Peruzzo}, \citenamefont {McClean}, \citenamefont {Shadbolt}, \citenamefont
  {Yung}, \citenamefont {Zhou}, \citenamefont {Love}, \citenamefont
  {Aspuru-Guzik},\ and\ \citenamefont {O'Brien}}]{Peruzzo2014}%
  \BibitemOpen
  \bibfield  {author} {\bibinfo {author} {\bibfnamefont {A.}~\bibnamefont
  {Peruzzo}}, \bibinfo {author} {\bibfnamefont {J.}~\bibnamefont {McClean}},
  \bibinfo {author} {\bibfnamefont {P.}~\bibnamefont {Shadbolt}}, \bibinfo
  {author} {\bibfnamefont {M.-H.}\ \bibnamefont {Yung}}, \bibinfo {author}
  {\bibfnamefont {X.-Q.}\ \bibnamefont {Zhou}}, \bibinfo {author}
  {\bibfnamefont {P.~J.}\ \bibnamefont {Love}}, \bibinfo {author}
  {\bibfnamefont {A.}~\bibnamefont {Aspuru-Guzik}}, \ and\ \bibinfo {author}
  {\bibfnamefont {J.~L.}\ \bibnamefont {O'Brien}},\ }\href {\doibase
  10.1038/ncomms5213} {\bibfield  {journal} {\bibinfo  {journal} {Nature
  Communications}\ }\textbf {\bibinfo {volume} {5}},\ \bibinfo {pages} {4213}
  (\bibinfo {year} {2014})}\BibitemShut {NoStop}%
\bibitem [{\citenamefont {Farhi}\ \emph {et~al.}(2014)\citenamefont {Farhi},
  \citenamefont {Goldstone},\ and\ \citenamefont {Gutmann}}]{farhi2014}%
  \BibitemOpen
  \bibfield  {author} {\bibinfo {author} {\bibfnamefont {E.}~\bibnamefont
  {Farhi}}, \bibinfo {author} {\bibfnamefont {J.}~\bibnamefont {Goldstone}}, \
  and\ \bibinfo {author} {\bibfnamefont {S.}~\bibnamefont {Gutmann}},\
  }\href@noop {} {\  (\bibinfo {year} {2014})},\ \Eprint
  {http://arxiv.org/abs/1411.4028} {arXiv:1411.4028 [quant-ph]} \BibitemShut
  {NoStop}%
\bibitem [{\citenamefont {McClean}\ \emph {et~al.}(2016)\citenamefont
  {McClean}, \citenamefont {Romero}, \citenamefont {Babbush},\ and\
  \citenamefont {Aspuru-Guzik}}]{McClean_2016}%
  \BibitemOpen
  \bibfield  {author} {\bibinfo {author} {\bibfnamefont {J.~R.}\ \bibnamefont
  {McClean}}, \bibinfo {author} {\bibfnamefont {J.}~\bibnamefont {Romero}},
  \bibinfo {author} {\bibfnamefont {R.}~\bibnamefont {Babbush}}, \ and\
  \bibinfo {author} {\bibfnamefont {A.}~\bibnamefont {Aspuru-Guzik}},\ }\href
  {\doibase 10.1088/1367-2630/18/2/023023} {\bibfield  {journal} {\bibinfo
  {journal} {New Journal of Physics}\ }\textbf {\bibinfo {volume} {18}},\
  \bibinfo {pages} {023023} (\bibinfo {year} {2016})}\BibitemShut {NoStop}%
\bibitem [{\citenamefont {Moll}\ \emph {et~al.}(2018)\citenamefont {Moll},
  \citenamefont {Barkoutsos}, \citenamefont {Bishop}, \citenamefont {Chow},
  \citenamefont {Cross}, \citenamefont {Egger}, \citenamefont {Filipp},
  \citenamefont {Fuhrer}, \citenamefont {Gambetta}, \citenamefont {Ganzhorn},\
  and\ \citenamefont {et~al.}}]{Moll_2018}%
  \BibitemOpen
  \bibfield  {author} {\bibinfo {author} {\bibfnamefont {N.}~\bibnamefont
  {Moll}}, \bibinfo {author} {\bibfnamefont {P.}~\bibnamefont {Barkoutsos}},
  \bibinfo {author} {\bibfnamefont {L.~S.}\ \bibnamefont {Bishop}}, \bibinfo
  {author} {\bibfnamefont {J.~M.}\ \bibnamefont {Chow}}, \bibinfo {author}
  {\bibfnamefont {A.}~\bibnamefont {Cross}}, \bibinfo {author} {\bibfnamefont
  {D.~J.}\ \bibnamefont {Egger}}, \bibinfo {author} {\bibfnamefont
  {S.}~\bibnamefont {Filipp}}, \bibinfo {author} {\bibfnamefont
  {A.}~\bibnamefont {Fuhrer}}, \bibinfo {author} {\bibfnamefont {J.~M.}\
  \bibnamefont {Gambetta}}, \bibinfo {author} {\bibfnamefont {M.}~\bibnamefont
  {Ganzhorn}}, \ and\ \bibinfo {author} {\bibnamefont {et~al.}},\ }\href
  {\doibase 10.1088/2058-9565/aab822} {\bibfield  {journal} {\bibinfo
  {journal} {Quantum Science and Technology}\ }\textbf {\bibinfo {volume}
  {3}},\ \bibinfo {pages} {030503} (\bibinfo {year} {2018})}\BibitemShut
  {NoStop}%
\bibitem [{\citenamefont {Cerezo}\ \emph
  {et~al.}(2020{\natexlab{a}})\citenamefont {Cerezo}, \citenamefont
  {Arrasmith}, \citenamefont {Babbush}, \citenamefont {Benjamin}, \citenamefont
  {Endo}, \citenamefont {Fujii}, \citenamefont {McClean}, \citenamefont
  {Mitarai}, \citenamefont {Yuan}, \citenamefont {Cincio},\ and\ \citenamefont
  {Coles}}]{cerezo2020variational}%
  \BibitemOpen
  \bibfield  {author} {\bibinfo {author} {\bibfnamefont {M.}~\bibnamefont
  {Cerezo}}, \bibinfo {author} {\bibfnamefont {A.}~\bibnamefont {Arrasmith}},
  \bibinfo {author} {\bibfnamefont {R.}~\bibnamefont {Babbush}}, \bibinfo
  {author} {\bibfnamefont {S.~C.}\ \bibnamefont {Benjamin}}, \bibinfo {author}
  {\bibfnamefont {S.}~\bibnamefont {Endo}}, \bibinfo {author} {\bibfnamefont
  {K.}~\bibnamefont {Fujii}}, \bibinfo {author} {\bibfnamefont {J.~R.}\
  \bibnamefont {McClean}}, \bibinfo {author} {\bibfnamefont {K.}~\bibnamefont
  {Mitarai}}, \bibinfo {author} {\bibfnamefont {X.}~\bibnamefont {Yuan}},
  \bibinfo {author} {\bibfnamefont {L.}~\bibnamefont {Cincio}}, \ and\ \bibinfo
  {author} {\bibfnamefont {P.~J.}\ \bibnamefont {Coles}},\ }\href@noop {} {\
  (\bibinfo {year} {2020}{\natexlab{a}})},\ \Eprint
  {http://arxiv.org/abs/2012.09265} {arXiv:2012.09265 [quant-ph]} \BibitemShut
  {NoStop}%
\bibitem [{\citenamefont {{Wolpert}}\ and\ \citenamefont
  {{Macready}}(1997)}]{Wolpert1997}%
  \BibitemOpen
  \bibfield  {author} {\bibinfo {author} {\bibfnamefont {D.~H.}\ \bibnamefont
  {{Wolpert}}}\ and\ \bibinfo {author} {\bibfnamefont {W.~G.}\ \bibnamefont
  {{Macready}}},\ }\href {\doibase 10.1109/4235.585893} {\bibfield  {journal}
  {\bibinfo  {journal} {IEEE Transactions on Evolutionary Computation}\
  }\textbf {\bibinfo {volume} {1}},\ \bibinfo {pages} {67} (\bibinfo {year}
  {1997})}\BibitemShut {NoStop}%
\bibitem [{\citenamefont {Poland}\ \emph {et~al.}(2020)\citenamefont {Poland},
  \citenamefont {Beer},\ and\ \citenamefont {Osborne}}]{poland2020no}%
  \BibitemOpen
  \bibfield  {author} {\bibinfo {author} {\bibfnamefont {K.}~\bibnamefont
  {Poland}}, \bibinfo {author} {\bibfnamefont {K.}~\bibnamefont {Beer}}, \ and\
  \bibinfo {author} {\bibfnamefont {T.~J.}\ \bibnamefont {Osborne}},\
  }\href@noop {} {\  (\bibinfo {year} {2020})},\ \Eprint
  {http://arxiv.org/abs/2003.14103} {arXiv:2003.14103 [quant-ph]} \BibitemShut
  {NoStop}%
\bibitem [{\citenamefont {Sharma}\ \emph {et~al.}(2020)\citenamefont {Sharma},
  \citenamefont {Cerezo}, \citenamefont {Holmes}, \citenamefont {Cincio},
  \citenamefont {Sornborger},\ and\ \citenamefont
  {Coles}}]{sharma2020reformulation}%
  \BibitemOpen
  \bibfield  {author} {\bibinfo {author} {\bibfnamefont {K.}~\bibnamefont
  {Sharma}}, \bibinfo {author} {\bibfnamefont {M.}~\bibnamefont {Cerezo}},
  \bibinfo {author} {\bibfnamefont {Z.}~\bibnamefont {Holmes}}, \bibinfo
  {author} {\bibfnamefont {L.}~\bibnamefont {Cincio}}, \bibinfo {author}
  {\bibfnamefont {A.}~\bibnamefont {Sornborger}}, \ and\ \bibinfo {author}
  {\bibfnamefont {P.~J.}\ \bibnamefont {Coles}},\ }\href@noop {} {\  (\bibinfo
  {year} {2020})},\ \Eprint {http://arxiv.org/abs/2007.04900} {arXiv:2007.04900
  [quant-ph]} \BibitemShut {NoStop}%
\bibitem [{\citenamefont {McClean}\ \emph {et~al.}(2018)\citenamefont
  {McClean}, \citenamefont {Boixo}, \citenamefont {Smelyanskiy}, \citenamefont
  {Babbush},\ and\ \citenamefont {Neven}}]{mcclean2018barren}%
  \BibitemOpen
  \bibfield  {author} {\bibinfo {author} {\bibfnamefont {J.~R.}\ \bibnamefont
  {McClean}}, \bibinfo {author} {\bibfnamefont {S.}~\bibnamefont {Boixo}},
  \bibinfo {author} {\bibfnamefont {V.~N.}\ \bibnamefont {Smelyanskiy}},
  \bibinfo {author} {\bibfnamefont {R.}~\bibnamefont {Babbush}}, \ and\
  \bibinfo {author} {\bibfnamefont {H.}~\bibnamefont {Neven}},\ }\href
  {\doibase 10.1038/s41467-018-07090-4} {\bibfield  {journal} {\bibinfo
  {journal} {Nature Communications}\ }\textbf {\bibinfo {volume} {9}},\
  \bibinfo {pages} {4812} (\bibinfo {year} {2018})}\BibitemShut {NoStop}%
\bibitem [{\citenamefont {Bravyi}\ \emph {et~al.}(2016)\citenamefont {Bravyi},
  \citenamefont {Smith},\ and\ \citenamefont {Smolin}}]{Bravyi2016}%
  \BibitemOpen
  \bibfield  {author} {\bibinfo {author} {\bibfnamefont {S.}~\bibnamefont
  {Bravyi}}, \bibinfo {author} {\bibfnamefont {G.}~\bibnamefont {Smith}}, \
  and\ \bibinfo {author} {\bibfnamefont {J.~A.}\ \bibnamefont {Smolin}},\
  }\href {\doibase 10.1103/physrevx.6.021043} {\bibfield  {journal} {\bibinfo
  {journal} {Physical Review X}\ }\textbf {\bibinfo {volume} {6}} (\bibinfo
  {year} {2016}),\ 10.1103/physrevx.6.021043}\BibitemShut {NoStop}%
\bibitem [{\citenamefont {Gr{\"o}tschel}\ and\ \citenamefont
  {Nemhauser}(1984)}]{Grotschel1984}%
  \BibitemOpen
  \bibfield  {author} {\bibinfo {author} {\bibfnamefont {M.}~\bibnamefont
  {Gr{\"o}tschel}}\ and\ \bibinfo {author} {\bibfnamefont {G.~L.}\ \bibnamefont
  {Nemhauser}},\ }\href {\doibase 10.1007/BF02591727} {\bibfield  {journal}
  {\bibinfo  {journal} {Mathematical Programming}\ }\textbf {\bibinfo {volume}
  {29}},\ \bibinfo {pages} {28} (\bibinfo {year} {1984})}\BibitemShut {NoStop}%
\bibitem [{\citenamefont {Peng}\ \emph {et~al.}(2020)\citenamefont {Peng},
  \citenamefont {Harrow}, \citenamefont {Ozols},\ and\ \citenamefont
  {Wu}}]{Peng2020}%
  \BibitemOpen
  \bibfield  {author} {\bibinfo {author} {\bibfnamefont {T.}~\bibnamefont
  {Peng}}, \bibinfo {author} {\bibfnamefont {A.~W.}\ \bibnamefont {Harrow}},
  \bibinfo {author} {\bibfnamefont {M.}~\bibnamefont {Ozols}}, \ and\ \bibinfo
  {author} {\bibfnamefont {X.}~\bibnamefont {Wu}},\ }\href {\doibase
  10.1103/PhysRevLett.125.150504} {\bibfield  {journal} {\bibinfo  {journal}
  {Phys. Rev. Lett.}\ }\textbf {\bibinfo {volume} {125}},\ \bibinfo {pages}
  {150504} (\bibinfo {year} {2020})}\BibitemShut {NoStop}%
\bibitem [{\citenamefont {Harrow}(2020)}]{harrow2020small}%
  \BibitemOpen
  \bibfield  {author} {\bibinfo {author} {\bibfnamefont {A.~W.}\ \bibnamefont
  {Harrow}},\ }\href@noop {} {\  (\bibinfo {year} {2020})},\ \Eprint
  {http://arxiv.org/abs/2004.00026} {arXiv:2004.00026 [quant-ph]} \BibitemShut
  {NoStop}%
\bibitem [{\citenamefont {Benedetti}\ \emph {et~al.}(2019)\citenamefont
  {Benedetti}, \citenamefont {Lloyd}, \citenamefont {Sack},\ and\ \citenamefont
  {Fiorentini}}]{Benedetti_2019}%
  \BibitemOpen
  \bibfield  {author} {\bibinfo {author} {\bibfnamefont {M.}~\bibnamefont
  {Benedetti}}, \bibinfo {author} {\bibfnamefont {E.}~\bibnamefont {Lloyd}},
  \bibinfo {author} {\bibfnamefont {S.}~\bibnamefont {Sack}}, \ and\ \bibinfo
  {author} {\bibfnamefont {M.}~\bibnamefont {Fiorentini}},\ }\href {\doibase
  10.1088/2058-9565/ab4eb5} {\bibfield  {journal} {\bibinfo  {journal} {Quantum
  Science and Technology}\ }\textbf {\bibinfo {volume} {4}},\ \bibinfo {pages}
  {043001} (\bibinfo {year} {2019})}\BibitemShut {NoStop}%
\bibitem [{\citenamefont {Schuld}\ and\ \citenamefont
  {Killoran}(2019)}]{Schuld2019}%
  \BibitemOpen
  \bibfield  {author} {\bibinfo {author} {\bibfnamefont {M.}~\bibnamefont
  {Schuld}}\ and\ \bibinfo {author} {\bibfnamefont {N.}~\bibnamefont
  {Killoran}},\ }\href {\doibase 10.1103/PhysRevLett.122.040504} {\bibfield
  {journal} {\bibinfo  {journal} {Phys. Rev. Lett.}\ }\textbf {\bibinfo
  {volume} {122}},\ \bibinfo {pages} {040504} (\bibinfo {year}
  {2019})}\BibitemShut {NoStop}%
\bibitem [{\citenamefont {Brandao}\ and\ \citenamefont
  {Svore}(2017)}]{brandao2017quantum}%
  \BibitemOpen
  \bibfield  {author} {\bibinfo {author} {\bibfnamefont {F.~G. S.~L.}\
  \bibnamefont {Brandao}}\ and\ \bibinfo {author} {\bibfnamefont
  {K.}~\bibnamefont {Svore}},\ }\href@noop {} {\  (\bibinfo {year} {2017})},\
  \Eprint {http://arxiv.org/abs/1609.05537} {arXiv:1609.05537 [quant-ph]}
  \BibitemShut {NoStop}%
\bibitem [{\citenamefont {Gondzio}(1998)}]{Gondzio1998}%
  \BibitemOpen
  \bibfield  {author} {\bibinfo {author} {\bibfnamefont {J.}~\bibnamefont
  {Gondzio}},\ }\href {\doibase 10.1007/bf02680554} {\bibfield  {journal}
  {\bibinfo  {journal} {Mathematical Programming}\ }\textbf {\bibinfo {volume}
  {83}},\ \bibinfo {pages} {125} (\bibinfo {year} {1998})}\BibitemShut
  {NoStop}%
\bibitem [{\citenamefont {Farhi}\ \emph {et~al.}(2017)\citenamefont {Farhi},
  \citenamefont {Goldstone}, \citenamefont {Gutmann},\ and\ \citenamefont
  {Neven}}]{farhi2017_q}%
  \BibitemOpen
  \bibfield  {author} {\bibinfo {author} {\bibfnamefont {E.}~\bibnamefont
  {Farhi}}, \bibinfo {author} {\bibfnamefont {J.}~\bibnamefont {Goldstone}},
  \bibinfo {author} {\bibfnamefont {S.}~\bibnamefont {Gutmann}}, \ and\
  \bibinfo {author} {\bibfnamefont {H.}~\bibnamefont {Neven}},\ }\href@noop {}
  {\  (\bibinfo {year} {2017})},\ \Eprint {http://arxiv.org/abs/1703.06199}
  {arXiv:1703.06199 [quant-ph]} \BibitemShut {NoStop}%
\bibitem [{\citenamefont {Egger}\ \emph {et~al.}(2020)\citenamefont {Egger},
  \citenamefont {Marecek},\ and\ \citenamefont
  {Woerner}}]{egger2020warmstarting}%
  \BibitemOpen
  \bibfield  {author} {\bibinfo {author} {\bibfnamefont {D.~J.}\ \bibnamefont
  {Egger}}, \bibinfo {author} {\bibfnamefont {J.}~\bibnamefont {Marecek}}, \
  and\ \bibinfo {author} {\bibfnamefont {S.}~\bibnamefont {Woerner}},\
  }\href@noop {} {\  (\bibinfo {year} {2020})},\ \Eprint
  {http://arxiv.org/abs/2009.10095} {arXiv:2009.10095 [quant-ph]} \BibitemShut
  {NoStop}%
\bibitem [{\citenamefont {Tate}\ \emph {et~al.}(2020)\citenamefont {Tate},
  \citenamefont {Farhadi}, \citenamefont {Herold}, \citenamefont {Mohler},\
  and\ \citenamefont {Gupta}}]{tate2020bridging}%
  \BibitemOpen
  \bibfield  {author} {\bibinfo {author} {\bibfnamefont {R.}~\bibnamefont
  {Tate}}, \bibinfo {author} {\bibfnamefont {M.}~\bibnamefont {Farhadi}},
  \bibinfo {author} {\bibfnamefont {C.}~\bibnamefont {Herold}}, \bibinfo
  {author} {\bibfnamefont {G.}~\bibnamefont {Mohler}}, \ and\ \bibinfo {author}
  {\bibfnamefont {S.}~\bibnamefont {Gupta}},\ }\href@noop {} {\  (\bibinfo
  {year} {2020})},\ \Eprint {http://arxiv.org/abs/2010.14021} {arXiv:2010.14021
  [quant-ph]} \BibitemShut {NoStop}%
\bibitem [{\citenamefont {H{\aa}stad}(2001{\natexlab{a}})}]{haastad2001some}%
  \BibitemOpen
  \bibfield  {author} {\bibinfo {author} {\bibfnamefont {J.}~\bibnamefont
  {H{\aa}stad}},\ }\href@noop {} {\bibfield  {journal} {\bibinfo  {journal}
  {Journal of the ACM (JACM)}\ }\textbf {\bibinfo {volume} {48}},\ \bibinfo
  {pages} {798} (\bibinfo {year} {2001}{\natexlab{a}})}\BibitemShut {NoStop}%
\bibitem [{\citenamefont {Berman}\ and\ \citenamefont
  {Karpinski}(1999)}]{berman1999some}%
  \BibitemOpen
  \bibfield  {author} {\bibinfo {author} {\bibfnamefont {P.}~\bibnamefont
  {Berman}}\ and\ \bibinfo {author} {\bibfnamefont {M.}~\bibnamefont
  {Karpinski}},\ }in\ \href@noop {} {\emph {\bibinfo {booktitle} {Automata,
  Languages and Programming}}}\ (\bibinfo  {publisher} {Springer Berlin
  Heidelberg},\ \bibinfo {address} {Berlin, Heidelberg},\ \bibinfo {year}
  {1999})\ pp.\ \bibinfo {pages} {200--209}\BibitemShut {NoStop}%
\bibitem [{\citenamefont {Khot}\ \emph {et~al.}(2007)\citenamefont {Khot},
  \citenamefont {Kindler}, \citenamefont {Mossel},\ and\ \citenamefont
  {O’Donnell}}]{khot2007optimal}%
  \BibitemOpen
  \bibfield  {author} {\bibinfo {author} {\bibfnamefont {S.}~\bibnamefont
  {Khot}}, \bibinfo {author} {\bibfnamefont {G.}~\bibnamefont {Kindler}},
  \bibinfo {author} {\bibfnamefont {E.}~\bibnamefont {Mossel}}, \ and\ \bibinfo
  {author} {\bibfnamefont {R.}~\bibnamefont {O’Donnell}},\ }\href {\doibase
  10.1137/S0097539705447372} {\bibfield  {journal} {\bibinfo  {journal} {SIAM
  J. Comput.}\ }\textbf {\bibinfo {volume} {37}},\ \bibinfo {pages} {319–357}
  (\bibinfo {year} {2007})}\BibitemShut {NoStop}%
\bibitem [{\citenamefont {Khot}(2002)}]{khotUG}%
  \BibitemOpen
  \bibfield  {author} {\bibinfo {author} {\bibfnamefont {S.}~\bibnamefont
  {Khot}}\ }(\bibinfo  {publisher} {Association for Computing Machinery},\
  \bibinfo {address} {New York, NY, USA},\ \bibinfo {year} {2002})\ p.\
  \bibinfo {pages} {767–775}\BibitemShut {NoStop}%
\bibitem [{\citenamefont {Garey}\ \emph {et~al.}(1976)\citenamefont {Garey},
  \citenamefont {Johnson},\ and\ \citenamefont {Stockmeyer}}]{GAREY1976}%
  \BibitemOpen
  \bibfield  {author} {\bibinfo {author} {\bibfnamefont {M.}~\bibnamefont
  {Garey}}, \bibinfo {author} {\bibfnamefont {D.}~\bibnamefont {Johnson}}, \
  and\ \bibinfo {author} {\bibfnamefont {L.}~\bibnamefont {Stockmeyer}},\
  }\href {\doibase https://doi.org/10.1016/0304-3975(76)90059-1} {\bibfield
  {journal} {\bibinfo  {journal} {Theoretical Computer Science}\ }\textbf
  {\bibinfo {volume} {1}},\ \bibinfo {pages} {237 } (\bibinfo {year}
  {1976})}\BibitemShut {NoStop}%
\bibitem [{\citenamefont {H{\aa}stad}(2001{\natexlab{b}})}]{Hstad2001}%
  \BibitemOpen
  \bibfield  {author} {\bibinfo {author} {\bibfnamefont {J.}~\bibnamefont
  {H{\aa}stad}},\ }\href {\doibase 10.1145/502090.502098} {\bibfield  {journal}
  {\bibinfo  {journal} {Journal of the {ACM}}\ }\textbf {\bibinfo {volume}
  {48}},\ \bibinfo {pages} {798} (\bibinfo {year}
  {2001}{\natexlab{b}})}\BibitemShut {NoStop}%
\bibitem [{\citenamefont {Zaslavsky}(2018{\natexlab{a}})}]{Zaslavsky2018}%
  \BibitemOpen
  \bibfield  {author} {\bibinfo {author} {\bibfnamefont {T.}~\bibnamefont
  {Zaslavsky}},\ }\href {\doibase 10.37236/29} {\bibfield  {journal} {\bibinfo
  {journal} {The Electronic Journal of Combinatorics}\ }\textbf {\bibinfo
  {volume} {1000}} (\bibinfo {year} {2018}{\natexlab{a}}),\
  10.37236/29}\BibitemShut {NoStop}%
\bibitem [{\citenamefont {Zaslavsky}(1982)}]{ZASLAVSKY1982}%
  \BibitemOpen
  \bibfield  {author} {\bibinfo {author} {\bibfnamefont {T.}~\bibnamefont
  {Zaslavsky}},\ }\href {\doibase https://doi.org/10.1016/0166-218X(82)90033-6}
  {\bibfield  {journal} {\bibinfo  {journal} {Discrete Applied Mathematics}\
  }\textbf {\bibinfo {volume} {4}},\ \bibinfo {pages} {47 } (\bibinfo {year}
  {1982})}\BibitemShut {NoStop}%
\bibitem [{\citenamefont {Harary}(1953)}]{harary1953}%
  \BibitemOpen
  \bibfield  {author} {\bibinfo {author} {\bibfnamefont {F.}~\bibnamefont
  {Harary}},\ }\href {\doibase 10.1307/mmj/1028989917} {\bibfield  {journal}
  {\bibinfo  {journal} {Michigan Mathematical Journal}\ }\textbf {\bibinfo
  {volume} {2}},\ \bibinfo {pages} {143 } (\bibinfo {year} {1953})}\BibitemShut
  {NoStop}%
\bibitem [{\citenamefont {Poljak}\ and\ \citenamefont
  {Tuza}(1993)}]{Poljak1993}%
  \BibitemOpen
  \bibfield  {author} {\bibinfo {author} {\bibfnamefont {S.}~\bibnamefont
  {Poljak}}\ and\ \bibinfo {author} {\bibfnamefont {Z.}~\bibnamefont {Tuza}},\
  }in\ \href@noop {} {\emph {\bibinfo {booktitle} {Combinatorial
  Optimization}}}\ (\bibinfo {year} {1993})\BibitemShut {NoStop}%
\bibitem [{\citenamefont {Harary}\ and\ \citenamefont
  {Kabell}(1980)}]{HARARY1980}%
  \BibitemOpen
  \bibfield  {author} {\bibinfo {author} {\bibfnamefont {F.}~\bibnamefont
  {Harary}}\ and\ \bibinfo {author} {\bibfnamefont {J.~A.}\ \bibnamefont
  {Kabell}},\ }\href {\doibase https://doi.org/10.1016/0165-4896(80)90010-4}
  {\bibfield  {journal} {\bibinfo  {journal} {Mathematical Social Sciences}\
  }\textbf {\bibinfo {volume} {1}},\ \bibinfo {pages} {131 } (\bibinfo {year}
  {1980})}\BibitemShut {NoStop}%
\bibitem [{\citenamefont
  {Zaslavsky}(2018{\natexlab{b}})}]{zaslavsky2018negative}%
  \BibitemOpen
  \bibfield  {author} {\bibinfo {author} {\bibfnamefont {T.}~\bibnamefont
  {Zaslavsky}},\ }\href@noop {} {\  (\bibinfo {year} {2018}{\natexlab{b}})},\
  \Eprint {http://arxiv.org/abs/1701.07963} {arXiv:1701.07963 [math.CO]}
  \BibitemShut {NoStop}%
\bibitem [{\citenamefont {Poljak}\ and\ \citenamefont
  {Turzík}(1986)}]{POLJAK1986}%
  \BibitemOpen
  \bibfield  {author} {\bibinfo {author} {\bibfnamefont {S.}~\bibnamefont
  {Poljak}}\ and\ \bibinfo {author} {\bibfnamefont {D.}~\bibnamefont
  {Turzík}},\ }\href {\doibase https://doi.org/10.1016/0012-365X(86)90192-5}
  {\bibfield  {journal} {\bibinfo  {journal} {Discrete Mathematics}\ }\textbf
  {\bibinfo {volume} {58}},\ \bibinfo {pages} {99 } (\bibinfo {year}
  {1986})}\BibitemShut {NoStop}%
\bibitem [{Note1()}]{Note1}%
  \BibitemOpen
  \bibinfo {note} {\protect \texttt {SIGNED MAXCUT} can be reduced to \protect
  \texttt {MAXCUT} by replacing every positive edge with two negative edges and
  a connectivity 2 vertex.}\BibitemShut {Stop}%
\bibitem [{\citenamefont {Ferris}\ and\ \citenamefont
  {Vidal}(2012)}]{ferris2012}%
  \BibitemOpen
  \bibfield  {author} {\bibinfo {author} {\bibfnamefont {A.~J.}\ \bibnamefont
  {Ferris}}\ and\ \bibinfo {author} {\bibfnamefont {G.}~\bibnamefont {Vidal}},\
  }\href {\doibase 10.1103/PhysRevB.85.165146} {\bibfield  {journal} {\bibinfo
  {journal} {Phys. Rev. B}\ }\textbf {\bibinfo {volume} {85}},\ \bibinfo
  {pages} {165146} (\bibinfo {year} {2012})}\BibitemShut {NoStop}%
\bibitem [{\citenamefont {{Shaydulin}}\ \emph {et~al.}(2019)\citenamefont
  {{Shaydulin}}, \citenamefont {{Safro}},\ and\ \citenamefont
  {{Larson}}}]{Shaydulin2019}%
  \BibitemOpen
  \bibfield  {author} {\bibinfo {author} {\bibfnamefont {R.}~\bibnamefont
  {{Shaydulin}}}, \bibinfo {author} {\bibfnamefont {I.}~\bibnamefont
  {{Safro}}}, \ and\ \bibinfo {author} {\bibfnamefont {J.}~\bibnamefont
  {{Larson}}},\ }in\ \href {\doibase 10.1109/HPEC.2019.8916288} {\emph
  {\bibinfo {booktitle} {2019 IEEE High Performance Extreme Computing
  Conference (HPEC)}}}\ (\bibinfo {year} {2019})\ pp.\ \bibinfo {pages}
  {1--8}\BibitemShut {NoStop}%
\bibitem [{\citenamefont {Cerezo}\ \emph
  {et~al.}(2020{\natexlab{b}})\citenamefont {Cerezo}, \citenamefont {Sone},
  \citenamefont {Volkoff}, \citenamefont {Cincio},\ and\ \citenamefont
  {Coles}}]{cerezo2020costfunctiondependent}%
  \BibitemOpen
  \bibfield  {author} {\bibinfo {author} {\bibfnamefont {M.}~\bibnamefont
  {Cerezo}}, \bibinfo {author} {\bibfnamefont {A.}~\bibnamefont {Sone}},
  \bibinfo {author} {\bibfnamefont {T.}~\bibnamefont {Volkoff}}, \bibinfo
  {author} {\bibfnamefont {L.}~\bibnamefont {Cincio}}, \ and\ \bibinfo {author}
  {\bibfnamefont {P.~J.}\ \bibnamefont {Coles}},\ }\href@noop {} {\  (\bibinfo
  {year} {2020}{\natexlab{b}})},\ \Eprint {http://arxiv.org/abs/2001.00550}
  {arXiv:2001.00550 [quant-ph]} \BibitemShut {NoStop}%
\bibitem [{\citenamefont {Bittel}\ and\ \citenamefont
  {Kliesch}(2021)}]{bittel2021}%
  \BibitemOpen
  \bibfield  {author} {\bibinfo {author} {\bibfnamefont {L.}~\bibnamefont
  {Bittel}}\ and\ \bibinfo {author} {\bibfnamefont {M.}~\bibnamefont
  {Kliesch}},\ }\href@noop {} {\  (\bibinfo {year} {2021})},\ \Eprint
  {http://arxiv.org/abs/2101.07267} {arXiv:2101.07267 [quant-ph]} \BibitemShut
  {NoStop}%
\bibitem [{\citenamefont {Crooks}(2018)}]{crooks2018performance}%
  \BibitemOpen
  \bibfield  {author} {\bibinfo {author} {\bibfnamefont {G.~E.}\ \bibnamefont
  {Crooks}},\ }\href@noop {} {\  (\bibinfo {year} {2018})},\ \Eprint
  {http://arxiv.org/abs/1811.08419} {arXiv:1811.08419 [quant-ph]} \BibitemShut
  {NoStop}%
\bibitem [{\citenamefont {Brandao}\ \emph {et~al.}(2018)\citenamefont
  {Brandao}, \citenamefont {Broughton}, \citenamefont {Farhi}, \citenamefont
  {Gutmann},\ and\ \citenamefont {Neven}}]{brandao2018}%
  \BibitemOpen
  \bibfield  {author} {\bibinfo {author} {\bibfnamefont {F.~G. S.~L.}\
  \bibnamefont {Brandao}}, \bibinfo {author} {\bibfnamefont {M.}~\bibnamefont
  {Broughton}}, \bibinfo {author} {\bibfnamefont {E.}~\bibnamefont {Farhi}},
  \bibinfo {author} {\bibfnamefont {S.}~\bibnamefont {Gutmann}}, \ and\
  \bibinfo {author} {\bibfnamefont {H.}~\bibnamefont {Neven}},\ }\href@noop {}
  {\  (\bibinfo {year} {2018})},\ \Eprint {http://arxiv.org/abs/1812.04170}
  {arXiv:1812.04170 [quant-ph]} \BibitemShut {NoStop}%
\bibitem [{\citenamefont {Zhou}\ \emph {et~al.}(2020)\citenamefont {Zhou},
  \citenamefont {Wang}, \citenamefont {Choi}, \citenamefont {Pichler},\ and\
  \citenamefont {Lukin}}]{Zhou_2020}%
  \BibitemOpen
  \bibfield  {author} {\bibinfo {author} {\bibfnamefont {L.}~\bibnamefont
  {Zhou}}, \bibinfo {author} {\bibfnamefont {S.-T.}\ \bibnamefont {Wang}},
  \bibinfo {author} {\bibfnamefont {S.}~\bibnamefont {Choi}}, \bibinfo {author}
  {\bibfnamefont {H.}~\bibnamefont {Pichler}}, \ and\ \bibinfo {author}
  {\bibfnamefont {M.~D.}\ \bibnamefont {Lukin}},\ }\href {\doibase
  10.1103/physrevx.10.021067} {\bibfield  {journal} {\bibinfo  {journal}
  {Physical Review X}\ }\textbf {\bibinfo {volume} {10}} (\bibinfo {year}
  {2020}),\ 10.1103/physrevx.10.021067}\BibitemShut {NoStop}%
\bibitem [{\citenamefont {Wurtz}\ and\ \citenamefont {Love}(2020)}]{wurtz2020}%
  \BibitemOpen
  \bibfield  {author} {\bibinfo {author} {\bibfnamefont {J.}~\bibnamefont
  {Wurtz}}\ and\ \bibinfo {author} {\bibfnamefont {P.~J.}\ \bibnamefont
  {Love}},\ }\href@noop {} {\  (\bibinfo {year} {2020})},\ \Eprint
  {http://arxiv.org/abs/2010.11209} {arXiv:2010.11209 [quant-ph]} \BibitemShut
  {NoStop}%
\bibitem [{Note2()}]{Note2}%
  \BibitemOpen
  \bibinfo {note} {Any classical algorithm can find solutions where the reduced
  balanced subgraph $\Gamma \protect \backslash e$ is connected and so will
  have an associated spanning tree by adding the following subroutine: for all
  vertices within some disconnected subgraph of $\Gamma \protect \backslash e$,
  swap the bipartition $X\Leftrightarrow Y$. This will satisfy all of the edges
  of the original graph between the two previously disconnected subgraphs,
  increasing the number of satisfied edges and making the reduced balanced
  graph connected.}\BibitemShut {Stop}%
\bibitem [{\citenamefont {Goemans}\ and\ \citenamefont
  {Williamson}(1995)}]{goemans1995}%
  \BibitemOpen
  \bibfield  {author} {\bibinfo {author} {\bibfnamefont {M.~X.}\ \bibnamefont
  {Goemans}}\ and\ \bibinfo {author} {\bibfnamefont {D.~P.}\ \bibnamefont
  {Williamson}},\ }\href {\doibase 10.1145/227683.227684} {\bibfield  {journal}
  {\bibinfo  {journal} {J. ACM}\ }\textbf {\bibinfo {volume} {42}},\ \bibinfo
  {pages} {1115–1145} (\bibinfo {year} {1995})}\BibitemShut {NoStop}%
\bibitem [{\citenamefont {Halperin}\ \emph {et~al.}(2004)\citenamefont
  {Halperin}, \citenamefont {Livnat},\ and\ \citenamefont
  {Zwick}}]{halperin2004max}%
  \BibitemOpen
  \bibfield  {author} {\bibinfo {author} {\bibfnamefont {E.}~\bibnamefont
  {Halperin}}, \bibinfo {author} {\bibfnamefont {D.}~\bibnamefont {Livnat}}, \
  and\ \bibinfo {author} {\bibfnamefont {U.}~\bibnamefont {Zwick}},\ }\href
  {\doibase https://doi.org/10.1016/j.jalgor.2004.06.001} {\bibfield  {journal}
  {\bibinfo  {journal} {Journal of Algorithms}\ }\textbf {\bibinfo {volume}
  {53}},\ \bibinfo {pages} {169} (\bibinfo {year} {2004})}\BibitemShut
  {NoStop}%
\end{thebibliography}%
	
\end{document}